\documentclass[12pt]{article}
\usepackage{amssymb,latexsym,epsfig,cite}

\begin{document}

\title{
Smarr's formula for black holes with non-linear electrodynamics}
\author{Nora Bret\'on \\
Departamento de F\'{\i}sica, Cinvestav-IPN, \\
Apdo. Postal 14-740, D.F., M\'exico.}


\maketitle

\begin{abstract}
 
It is known that for nonlinear electrodynamics the First Law of Black Hole
Mechanics holds, however the Smarr's formula for the total mass does not.
In this contribution we discuss the point and determine the corresponding
expressions for the Bardeen black hole solution that represents a
nonlinear magnetic monopole. The same is done for the regular black hole
solution derived by Ay\'on-Beato and Garc\'{\i}a \cite{ABG-PRL}, showing
that in the case that variations of the electric charge are involved, the
Smarr's formula does not longer is valid. \footnote{\it This contribution
is devoted to Profr. Alberto Garc\'{\i}a who introduced me into the
General Relativity world.} \end{abstract}

\section{Introduction}

The Reissner-Nordstr\"om (RN) solution is the unique static and
asymptotically flat solution to the Einstein-Maxwell equations for
spherical symmetry. It represents a black hole characterized by its mass
and electric charge. The geometry of the RN black hole is singular at the
origin of the radial coordinate, then it has been a subject of research
the construction of its regular generalizations. A good candidate for the
source term of the Einstein equations is the (classical) stress-energy
tensor of nonlinear electrodynamics. The recent renewal of interest in
nonlinear electrodynamics has to do also with the fact that such theories
arose as limiting cases of certain formulations of string theory.
   
Nonlinear or logarithmic electromagnetic Lagrangians coupled to gravity
have been studied in an attempt to remove some of the singularities
associated with charged black holes. The exterior of such black holes is,
at large distances, the same as the usual (RN) black holes of
Einstein-Maxwell theory.  Close to the black hole, however, things may be
very different. Quantities defined at the horizon of such black holes have
been useful in obtaining a more accurate description of the physics near
the black hole. In this regard, the concept of {\it isolated horizon} has
been used to provide a full Hamiltonian treatment of black holes
\cite{ACS}. This framework has been applied succesfully not only to
Einstein-Maxwell theory but also in more general cases like non-Abelian
gauge theories \cite{ulises}. At this point it is relevant to question the
role of the Laws of Black Hole Mechanics (BHM) in this description, i. e.
in situations more general than Einstein-Maxwell fields.
  
The zeroth and first laws of BHM refer to equilibrium situations and small
departures therefrom. First Law of BHM is an identity relating the changes
in mass, angular momentum and horizon area of a stationary black hole when
it is perturbed. The variation applies for perturbations from one
stationary axisymmetric solution of Einstein equations to another;
moreover, it has been shown that the validity of this law depends only on
very general properties of the field equations \cite{Wald}. For the
horizon mass $M_{\Delta}$, the first law when static spherically symmetric
solutions are considered \cite{ulises}, is

\begin{equation}
\delta M_{\Delta}= \frac{\kappa}{8 \pi} \delta{a_{\Delta}} 
+ \Phi_{\Delta} \delta Q_{\Delta}, 
\label{first}
\end{equation}
where $\kappa$ is the surface gravity at the horizon, $a$ is the area of
the horizon, $Q$ is the electric charge and $\Phi$ is the
electric potential; the subindex $\Delta$ indicates that the quantity is
evaluated at the horizon of the black hole.
  
On the other side, the total mass is given by the Smarr's formula

\begin{equation}
M_{\Delta}= \frac{\kappa a_{\Delta}}{4 \pi} + 
\Phi_{\Delta} Q_{\Delta} .
\label{smarr}
\end{equation}
 
In the case of Einstein-Maxwell theory, it is possible to deduce one, Eq.
(\ref{first}), directly from the other, Eq. (\ref{smarr}), using the
homogeneity of the mass as a function of $\sqrt{a}$ and $Q$. In the work
by Ashtekar, Corichi and Sudarsky \cite{ACS} the first law of BHM, for
quantities defined only at the horizon, arises naturally as part of the
requirements for a consistent Hamiltonian formulation. In the case of
non-linear electrodynamics, however, one no longer has homogeneity of the
mass function and a priori one has no reason to expect that either of them
holds.
 
Previous work on this line includes the derivation of the first law of
black hole physics for some nonlinear matter models \cite{Heusler}. D. A.
Rasheed \cite{Rasheed} studied the Zeroth and First Laws of BHM in the
context of non-linear electrodynamics coupled to gravity. In this case,
the Zeroth Law, which states that the surface gravity of a stationary
black hole is constant over the event horizon, is shown to hold even if
the Dominant Energy Condition \cite{Hawk} is violated. In the same paper,
it is found that the usual First Law (the general mass variation formula)
holds true for the case of non-linear electrodynamics but the formula for
the total mass, known as Smarr's formula, does not.
    
However, we can propose the form that must have a Smarr-type formula for
the horizon mass in order to be consistent with the variations expressed
by the first law of BHM that indeed holds,

\begin{equation}
M_{\Delta}= \frac{\kappa a_{\Delta}}{4 \pi} + 
\Phi_{\Delta} Q_{\Delta}  + V(a_{\Delta}, Q_{\Delta}, P_{\Delta}),
\label{hormass0}
\end{equation}
where $V$ is a so far undetermined potential that depends on the horizon
parameters, $a_{\Delta}, Q_{\Delta}, P_{\Delta}$ and also of the coupling
constants of the theory. In the variational principle this term plays no
role, however in the Hamiltonian description it becomes essential.
 
Note that in the first law, Eq. (\ref{first}), only variations of the
electric charge are involved, and not variations of the magnetic charge.
On the other hand, the horizon mass, Eq. (\ref{hormass0}) might depend on
$P_{\Delta}$ through $V$.
 
The equations to determine the potential $V(a_{\Delta},
Q_{\Delta}, P_{\Delta})$ arise from the condition that the first law
holds and demanding consistency between Eq.(\ref{first}) and the
variations of Eq.(\ref{hormass0}), 

\begin{eqnarray}
a_{\Delta} \frac{\partial \beta}{\partial a_{\Delta}}+
8 \pi r_{\Delta} Q_{\Delta}  \frac{\partial \Phi}{\partial
a_{\Delta}} + 8 \pi r_{\Delta}  \frac{\partial V}{\partial 
a_{\Delta}}&&=0,  \nonumber\\
\frac{r_{\Delta}}{2} \frac{\partial \beta}{\partial Q_{\Delta}}+
Q_{\Delta}  \frac{\partial \Phi}{\partial Q_{\Delta}} 
+ \frac{\partial V}{\partial Q_{\Delta}}&&=0,
\label{Veqs}
\end{eqnarray} 
where $\beta = 1-2m'(r)$, $a_{\Delta}=4 \pi r_{\Delta}^2$
and $r_{\Delta}$ is the radius of the horizon.

The condition of consistency determines the set of parameters that can
vary independently, in this case, the magnetic charge becomes a function
of the area and electric charge, $P_{\Delta}= P_{\Delta}(r_{\Delta},
Q_{\Delta})$.
 
In what follows we shall determine the horizon mass in agreement with the
first law of BHM for the Bardeen black hole and then for the regular black
hole of ABG, both solutions of Einstein equations coupled with nonlinear
electrodynamics.

We shall consider non-linear electrodynamics governed by
an action of the form

\begin{equation}
S = \int{d^4x \sqrt{-g} \{ R (16\pi)^{-1}+ {\cal L} \} } ,
\end{equation}
where $R$ denotes the scalar curvature, $g:= {\rm det} \vert g_{\mu \nu}
\vert$ and ${\cal L}$, the electromagnetic part, is assumed to depend in
nonlinear way on the invariants of the field strength tensor $F_{\mu
\nu}$. As we mentioned above, this kind of fields have been studied with
the aim of avoiding the singularities of black holes and other systems
\cite{NLEbh}. Succesful advances on this line were the proposed nonlinear
electrodynamics theory by Born and Infeld \cite{Born} which in fact
succeeded in avoiding the electric field singularity at the charge
position.  Born-Infeld theory coupled to gravity were studied by Hoffmann
and Infeld \cite{HI}.  Also we must mention the pioneering work done by A.
Garc\'{\i}a as colaborator of Profr. J. Pleba\~nski, in studying the
problem of non-linear electrodynamics for the type D solutions of the
Einstein-Born-Infeld coupled equations \cite{GSP1}.
     
In the search for a regular black hole with nonlinear electrodynamics,
there exists a no go theorem \cite{Bronnikov1} which states that for
Lagrangians depending on the invariant of the electromagnetic field,
${\cal{L}}(F), F=F_{\mu \nu}F^{\mu \nu}$, with the Maxwell weak-field
limit, there are no spherically symmetric static black hole solutions with
a regular center. However, regular solutions with only {\it magnetic}
charge may exist \cite{Bronnikov2}. It is not excluded neither the
possibility of regular solutions corresponding to Lagrangians depending on
both invariants of the electromagnetic field, ${\cal{L}}(F,Q), F=F_{\mu
\nu}F^{\mu \nu}, Q= \tilde{F}_{\mu \nu}F^{\mu \nu}$, with $\tilde{F}_{\mu
\nu}$ being the dual of ${F}_{\mu \nu}$. Recently, several solutions
corresponding to regular black holes with nonlinear electrodynamics with
Lagrangians of the form ${\cal{L}}(F)$,have been derived \cite{ABG}.
Moreover, it has been of interest the interpretation given to the model of
Bardeen for a regular black hole, as corresponding to a self-gravitating
magnetic monopole. The Bardeen model was proposed some years ago as a
regular black hole, however, only recently it has been shown
\cite{ABGBardeen} that it is an exact solution of the Einstein equations
coupled to a kind of nonlinear electrodynamics characterized by the
Lagrangian

\begin{equation}
{\cal L}(F)= \frac{2}{2sg^2}(\frac{2g^2F}{1+ \sqrt{2g^2F}})^{5/2},
\end{equation} 
The corresponding energy momentum tensor fulfills the weak energy
condition and is regular everywhere. For a spherically symmetric space,
the corresponding metric is given by

\begin{eqnarray}
ds^2&=& - \psi_B dt^2+\psi_B^{-1}dr^2+r^2(d \theta^2 + \sin^2{\theta}d
\phi^2), \\ 
\psi_B&=& 1- \frac{2m(r)}{r}= 1- \frac{2mr^2}{(r^2+g^2)^{3/2}},
\label{metrfunc}
\end{eqnarray}

This solution is a self-gravitating magnetic monopole with charge
$g$. The solution is regular everywhere, although the invariants of
the electromagnetic field exhibit the usual singular behaviour of
magnetic monopoles, $F=F^{\mu \nu}_{\mu \nu}= g^2/2r^4$.
The asymptotic behaviour of the solution is

\begin{equation}
\psi_B=g_{tt} \approx 1-\frac{2m}{r}+\frac{3mg^2}{r^3},
\end{equation}
it is this behavior at infinity, in which the constant $g$ vanishes as
$1/r^3$, and not as a Coulombian term ($1/r^2$), that allows to interpret
the constant $g$ as a magnetic charge.
   
The horizons are given by the roots of the equation $r=2m(r)$. In this
case these roots are not as easy to calculate as for RN. Hence the
conditions which restrict the parameters in order that the solution
corresponds to regions where $\psi_B \ge 0$ are more difficult to find.
The Bardeen solution does not involve electric charge, then the horizon
mass depends only on the area of the horizon,

\begin{equation}
M_{\Delta}= \frac{1}{8 \pi} \int{\kappa da}= \int{(1-m')dr},
\label{hormass2}
\end{equation}
the condition that the horizon mass be positive, from Eq.
(\ref{hormass2}), amounts to $m(r) \le r$, this condition also guarantees
that $g_{tt} \ge 0$. Using the expression for $g_{tt}$ it amounts to
$(r^2+g^2)^3 \ge 4m^2r^4$. In this case when $g^2=\frac{16}{27}m^2$ the
two horizons that could be present shrink into a single one, being this
value of $g$ the corresponding to the extreme black hole; for $g^2<
\frac{16}{27}m^2$ there exist both inner and event horizon.

The potential $V$ for the Smarr-type formula, Eq. (\ref{hormass0}), for
the Bardeen black hole turns out to be, undetermined until an integration
constant which we have put zero,

\begin{equation}
V= mr^3 \frac{2g^2-r^2}{(g^2+r^2)^{\frac{3}{2}}},
\end{equation}

Substituting $V$ in the Smarr-type formula one obtains the
horizon mass

\begin{equation}
M_{\Delta}= \frac{r}{2}-\frac{mr^3}{(r^2+g^2)^{\frac{3}{2}}}
\label{hormass3}
\end{equation}

This value for the horizon mass coincides with the one
determined by integrating the first law, Eq. (\ref{hormass2}).

\begin{figure}\centering
\epsfig{file=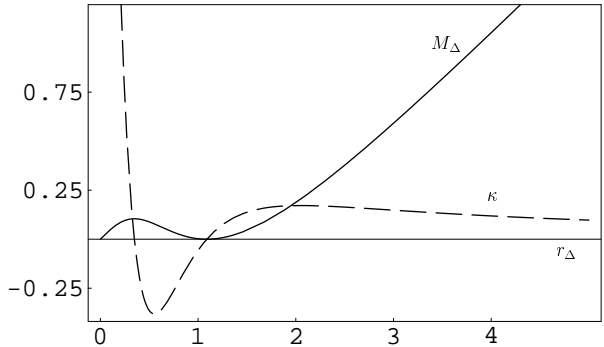, width=8cm}
\caption{
Horizon mass $M_{\Delta}$ and surface gravity $\kappa$, as functions of
the horizon radious, for the extreme Bardeen black hole, in this case the
magnetic charge has the value $g^2=16m^2/27$.}
\label{Bbhextreme}
\end{figure}

\begin{figure}\centering
\epsfig{file=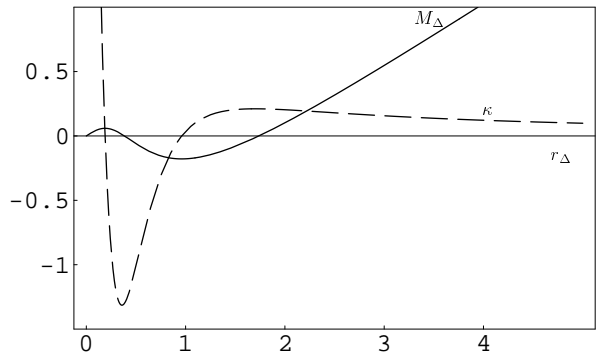, width=8cm}
\caption{
Horizon mass $M_{\Delta}$ and surface gravity $\kappa$, for a Bardeen
black hole with magnetic charge $g^2=18m^2/27$, in this case the black
hole present an inner and an event horizon. $M_{\Delta}$ has meaning only
in the range for which $M_{\Delta} \ge 0$ .}
\label{Bbh}
\end{figure}

In Fig. \ref{Bbhextreme} we have depicted both the horizon mass
$M_{\Delta}$ and the surface gravity
$\kappa=(1-6mr_{\Delta}^2g^2(r_{\Delta}^2+g^2)^{\frac{5}{2}})/2r_{\Delta}$,
for the extreme case. In Fig. \ref{Bbh} are shown the horizon mass and the
surface gravity in the case when inner and event horizon appear. In both
figures there are regions where the horizon mass and the surface gravity
are negative, these are regions inner to the event horizon, so they have
no meaning as the formalism is valid for the regions $M_{\Delta} \ge 0$.  
As we pointed out above, the condition for the positiveness of
$M_{\Delta}$ is the same for $g_{tt}$. Furthermore, we remind that for the
regions where $g_{tt} \le 0$ the signature of the metric changes, as does
the character of the Killing vectors, in such a manner that it is a
spatially homogeneous region that is not static; then the situation does
not correspond to an always incresing area of the horizon, but on the
contrary, as we penetrate the spatially homogeneous region, the area
decreases reaching a minimum and then increasing again. The Bardeen black
hole turns out to be stable with respects to arbitrary linear fluctuations
of the metric and electromagnetic field \cite{Clau}.
 
Note that the horizon mass of the Bardeen black hole involved dependence
only on the horizon area, since the magnetic charge is not considered as a
varying parameter of the horizon. So far we have shown the agreement
between the horizon mass calculated with the first law of BHM and when it
is determined by adding the apropriate potential to a Smarr-type formula.

However, things are not so easy when we consider situations involving
electric charge in nonlinear electrodynamics. This is the case of the
regular black hole derived by Ay\'on-Beato and Garc\'{\i}a (ABG)
\cite{ABG-PRL}, that we have to take into account in the corresponding
potential $V$, the term of the variation of the electric charge. The
metric of the ABG black hole is spherically symmetric, given by

\begin{eqnarray}
ds^2&=& - \psi_{ABG} dt^2+\psi_{ABG}^{-1}dr^2+r^2(d \theta^2 +
\sin^2{\theta}d
\phi^2), \\
\psi_{ABG}&=&  1- \frac{2mr^2}{(r^2+q^2)^{3/2}}+
\frac{q^2r^4}{(r^2+q^2)^2},
\label{metrfuncABG}
\end{eqnarray}
Asymptotically this solution behaves as a RN.
The ABG line element is not a solution of the standard nonlinear
electrodynamics and the effective geometry (i.e. the geometry affecting
the photons of the nonlinear theory) is singular. This regular black hole
is an exact solution of Einstein equations coupled with a Lagrangian
matter of the form 

\begin{equation}
{\cal{L}} =P \frac{1-8
\sqrt{-2q^2 P}-6q^2P}{(1+\sqrt{-2q^2P})^4}-\frac{3}{4q^2s}
\frac{(-2q^2P)^{\frac{5}{4}}(3-2\sqrt{-2q^2P})}{(1+\sqrt{-2q^2P})^{\frac{7}{2}}}
\label{NLE-ABG}
\end{equation}
where $P$ is the invariant of the electromagnetic field tensor $P_{\mu
\nu}$ and $s= \vert q \vert /2m$. The no go theorem about a regular static
spherically solution with electric charge can be eluded reinterpreting the
ABG solution as describing a magnetically charged regular solution of the
coupled equations of nonlinear electrodynamics of Eq. (\ref{NLE-ABG}) and
gravitation with much more regular behaviour of the effective geometry.
 
The Smarr-type formula Eq. (\ref{hormass0}), with the potential $V$
determined from Eqs. (\ref{Veqs}) amounts to

\begin{eqnarray}
M_{\Delta}=&& \frac{r}{2}- \frac{3mr}{2
\sqrt{r^2+q^2}}-\frac{r^5}{4(r^2+q^2)^2}+
\frac{3m}{2} \ln [\frac{r+\sqrt{r^2+q^2}}{q}] \nonumber\\
&& + \frac{3mr^3q^2}{2(r^2+q^2)^{\frac{5}{2}}}-
\frac{r^3q^4}{(r^2+q^2)^3},
\end{eqnarray}
   
While the expression for the horizon mass determined from the first law of
BHM, Eq. (\ref{first}), taking into account the presence of the electric
charge, is

\begin{eqnarray}
M=&& \frac{1}{8 \pi} \int{\kappa da} + \int{\phi dq}, \nonumber\\
= && \frac{r}{2}- \frac{3mr}{2
\sqrt{r^2+q^2}}-\frac{r^5}{4(r^2+q^2)^2}+
\frac{3m}{2} \ln [\frac{r+\sqrt{r^2+q^2}}{q}] \nonumber\\
&& -\frac{3mr^3}{2(r^2+q^2)^{\frac{3}{2}}}+
\frac{r^3q^2}{2(r^2+q^2)^2},
\end{eqnarray}  

\begin{figure}\centering
\epsfig{file=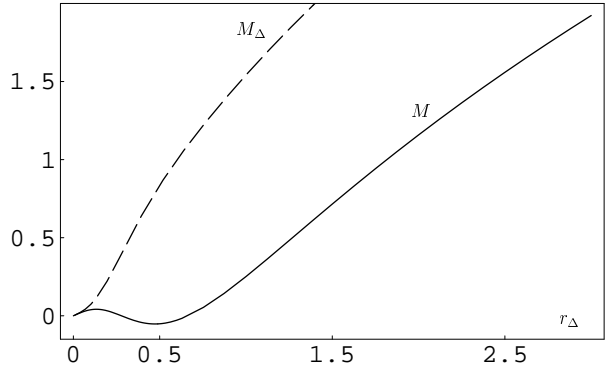, width=8cm}
\caption{
The horizon mass for the Ay\'on-Beato-Garc\'{\i}a black hole is displayed.
The discontinuous curve corresponds to $M_{\Delta}$ calculated using a
Smarr-type formula while the continuous one, also for the horizon mass
$M$, was determined by using the first law of BHM .}
\label{ABGbh}
\end{figure}

The two expressions differ in the last two terms, the mismatch is showed
in Fig. \ref{ABGbh} for the horizon mass. Remains as an open problem the
reason why the potential $V$ determined in agreement with the first law of
BHM can not give the appropriate dependence for the terms corresponding to
the electric charge. Eqs. (\ref{Veqs}) do not describe in a feasible form
the potential $V$ in situations where nonlinear electromagnetic fields are
present. It might be that the dependence of $V$ on the charge is of a
nonlinear nature that can not be approached with Eqs.(\ref{Veqs}).

If in contrast, we reinterpret the ABG solution as describing a
magnetically charged regular solution, then the potential does not depend
on the magnetic charge, but solely on the horizon area. In this case the
agreement between the two procedures to calculate the horizon mass is
held and the corresponding expression is

\begin{equation}
M_{\Delta}= \frac{r}{2}- \frac{mr^3}{
(r^2+q^2)^{\frac{3}{2}}}+\frac{r^3q^2}{2(r^2+q^2)^2}.
\end{equation}
  
In this contribution we have illustrated that in cases involving nonlinear
electromagnetic fields, the horizon mass calculated with a Smarr type
formula that is consistent with the fist law of BHM, is feasible only for
the magnetic sector of the solutions. If the variation of electric charge
is taken into account in the potential $V$ of the Smarr formula, the
mentioned consistency does not longer hold. The cases we addressed in this
regard were the Bardeen magnetic monopole and the regular black hole
derived by Ay\'on-Beato and Garc\'{\i}a.

\end{document}